\documentclass[twocolumn, pra, showpacs]{revtex4}
\usepackage{amssymb}
\usepackage{graphicx}
\usepackage{dcolumn}
\usepackage{bm}
\usepackage{amsmath}
\usepackage[dvipdfm,colorlinks,linkcolor=red,citecolor=blue]{hyperref}

\begin{document}

\title{Spin-orbit-angular-momentum coupling in a spin-1 Bose-Einstein condensate}
\author{Li Chen$^{1}$}
\author{Han Pu$^{2,3}$}
\email{hpu@rice.edu}
\author{Yunbo Zhang$^1$}
\email{ybzhang@sxu.edu.cn}
\affiliation{$^1${Institute of Theoretical Physics, Shanxi University, Taiyuan, Shanxi 030006, P. R.
China}\\
$^2${Department of Physics and Astronomy, and Rice Center for Quantum
Materials, Rice University, Houston, TX 77005, USA}\\
$^3${Center for Cold Atom Physics, Chinese Academy of Sciences, Wuhan
430071, China}}

\begin{abstract}
We propose a simple model with spin and orbit angular momentum coupling in a
spin-1 Bose-Einstein condensate, where three internal atomic states are Raman
coupled by a pair of co-propagating Laguerre-Gaussian beams. The resulting Raman transition imposes a transfer of orbital angular momentum between photons and the condensate in a spin-dependent way. Focusing on a regime where the single-particle ground state is nearly three-fold degenerate, we show that the weak interatomic interaction in the condensate produces a rich phase diagram, and that a many-body Rabi oscillation between two quantum phases can be induced by a sudden quench of the quadratic Zeeman shift. We carried out our calculations using both a variational method and a full numerical method, and found excellent agreement.
\end{abstract}

\maketitle

\section{INTRODUCTION}

A promising platform to quantum simulate such novel phenomena of condensed matter
physics as topological insulators \cite{Hasan} and superconductors
\cite{Qi} is the spin-orbit coupled cold atomic systems \cite{Dalibard,Goldman,Zhai} which have
drawn great attention in recent years. Raman
dressed coupling between atomic pseudo-spin and its linear momentum was first realized by
Lin and co-workers \cite{Lin} in a two-component (spin-1/2) $^{87}$Rb condensate, and were soon generalized
to spin-1/2 degenerate Fermi gases of $^{40}$K \cite{Wang} and $^{6}$Li
\cite{Cheuk}. In this scheme, photon recoil associated with the Raman transition --- facilitated by two counter-propagating laser beams --- changes the center-of-mass momentum of the atom when it jumps from one spin state to the other. Very recently, the same spin-orbit coupling (SOC) scheme was achieved in a spin-1 condensate \cite{Campbell}. In general, the physics becomes richer in larger spin systems \cite{Lan,Natu,Sun2015} simply because more spin states are involved and more control knobs can be utilized.

When the two laser beams that induce the Raman transition are made to co-propagating, but possess different orbital angular momenta (e.g., in the form of Laguerre-Gaussian, or LG, beams) \cite{Allen}, the Raman transition will be accompanied by a transfer of the orbital angular momentum (OAM), instead of the linear momentum, to the atom. This situation has been achieved in experiment where this transfer of OAM from photon to the atom was exploited to create spin textures in a spinor condensate \cite{Wright2008,Wright2009}. Several recent theoretical proposals also explored this effect to realize spin-orbit-angular-momentum
(SOAM) coupling in spin-1/2 condensate \cite{DeMarco,Sun,Qu,Hu}, where interesting
quantum states such as the half Skyrmion, the meron pair and the
annular striped phase, are predicted to exist.

In this paper, we provide our study of the SOAM coupling in a weakly interacting spin-1 condensate. We show that the interplay between the SOAM coupling and the quadratic Zeeman shift produces a rich phase diagram, and a coherent oscillation between two different many-body quantum phases can be induced by quenching the quadratic Zeeman term.

Our paper is organized as follows. In Sec. II, we present the model and discuss the single-particle properties, particularly the single-particle energy spectrum, of the system. In Sec. III, we focus on the properties of a spin-1 condensate. Both the ground state properties and the quench dynamics will be presented. In Sec. IV, we provide a brief summary.

\section{HAMILTONIAN and single-particle physics}

\subsection{Model and single-particle Hamiltonian}

\begin{figure}[t]
\includegraphics[width=0.5\textwidth]{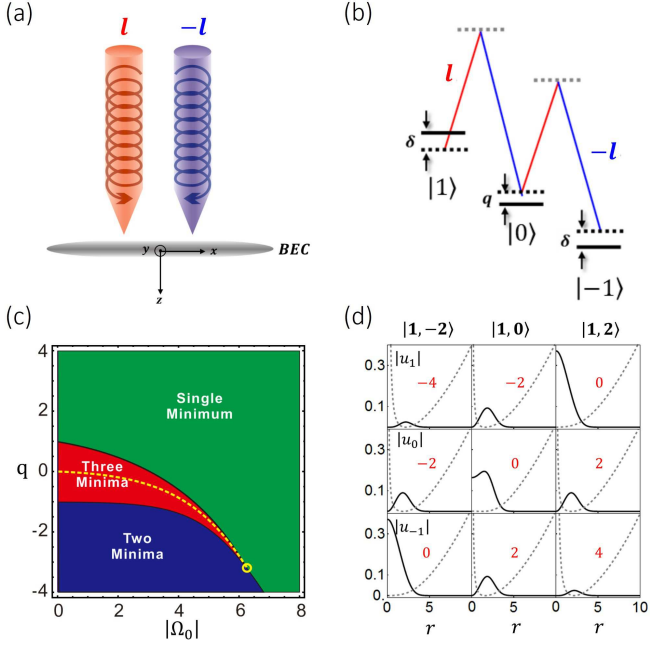}
\caption{(Color Online) (a) A schematic picture showing two LG beams with opposite OAM co-propagating along the $z$-axis shine on a condensate. (b) Atomic energy level
structure. (c) The lowest energy band ($n=1$) in the single-particle spectrum possesses different number of minima, which yield this  phase diagram in the $\left\vert \Omega _{0}\right\vert$-$q$ space. On the yellow dashed line, the single-particle ground state
is three-fold degenerate. (d) Solid lines represent the three degenerate ground state wave functions $|\psi_m|=|u_m|$ ($m=-1$, 0, and 1). Here $\Omega=-4$ and $q = -0.817$. The dashed lines represent the effective potential $V_m$ experienced by different spin states. The red numbers in the figure represent the OAM quantum number is the lab frame, which equal to $\widetilde{l}_z \mp 2m l$, and we take $l=1$ in all our calculations. All quantities plotted in the figures throughout the paper are expressed in a dimensionless unit system with $\hbar =M=\omega=1$.}
\label{Fig1}
\end{figure}
Our theoretical model is similar to the one presented in an earlier work for spin-1/2 system \cite{DeMarco}, where two LG beams co-propogating along the $z$-axis carrying opposite angular momenta ($%
l\hbar $ and $-l\hbar $) shine on a harmonically-trapped
condensate, as schematically shown in Fig.~\ref{Fig1}(a). The relevant atomic energy levels are shown in Fig.~\ref{Fig1}(b). The two laser beams induce Raman transition among the three atomic hyperfine spin states (denoted as $\left\vert 1\right\rangle
,\left\vert 0\right\rangle $ and $\left\vert -1\right\rangle $), which form a spin-1 system. In experiment, one may choose, for example, the three
Zeeman levels in $^{87}$Rb, e.g. $\left\vert
F=2,m_{F}=-2\right\rangle $, $\left\vert F=2,m_{F}=0\right\rangle $ and $\left\vert
F=2,m_{F}=2\right\rangle $ in the $F=2$ ground hyperfine manifold \cite{Wright2008}. We assume that the single-photon detuning is sufficiently large such that the atomic electronically excited states can be adiabatically eliminated. For simplicity, we also assume that the
two LG beams have identical spatial profile and light intensity. Finally, we assume that the atom is tightly confined along the $z$-axis which results in a quasi-two-dimensional geometry. Under the rotating wave approximation \cite{Scully}, by adopting a unit system such that $\hbar=M=\omega=1$ with $M$ and $\omega$ being respectively the atomic mass and harmonic trap frequency, the
dimensionless single-particle Hamiltonian can be written in polar coordinates $(r, \phi)$
in the following form
\begin{eqnarray}
H_{0} &=&-\frac{1}{2}\nabla ^{2}+\frac{1}{2}r^{2}+2\Omega _{R}\left(
r\right) +\delta \hat{S}_{z}+\left( q-\Omega _{R}\right) \hat{S}_{z}^{2}  \notag \\
&&+\sqrt{2}\Omega _{R}\left[ \cos \left( 2l\phi \right) \hat{S}_{x}+\sin \left(
2l\phi \right) \hat{S}_{y}\right] \,.  \label{C1}
\end{eqnarray}%
Here $\hat{\bf S} = (\hat{S}_{x},\,\hat{S}_y,\,\hat{S}_{z})$ are the spin-1 angular momentum matrices, $\Omega _{R}\left( r\right) =2\Omega _{0}\left( \frac{r}{w}\right)
^{2l}\left[ L_{k}^{l}\left( \frac{2r^{2}}{w^{2}}\right) e^{-r^{2}/w^{2}}%
\right] ^{2}$ represents both the Raman coupling strength and the AC Stark shift, where $\Omega
_{0}$ is a constant proportion to the overall light intensity, $w$ characterizes the beam width, and $L_{k}^{l}$ is the generalized Laguerre polynomials
with azimuthal index $l$ determining the optical OAM and the radial index $k$
describing the radial intensity distribution of the LG beams \cite{Allen}. Finally, the parameters $%
\delta $ and $q$ denote the effective linear and quadratic Zeeman shifts, respectively. Physically, $\delta$ is related to the two-photon Raman detuning, and $q$ can be tuned by either an external magnetic field or a microwave field and can be either positive or negative \cite%
{Gerbier}.

Next we introduce a rotating frame which is related to the lab frame by a unitary transformation with the corresponding unitary operator $U=e^{2il\phi \hat{S}_{z}}$. Under this unitary transformation, the atomic states $\Psi = (\psi_1,\, \psi_0,\, \psi_{-1})^T$ are transformed to $\widetilde{\Psi} = U\Psi =(e^{2il\phi }\psi _{1},\, \psi_0,\, e^{-2il\phi }\psi _{-1})^T $, and the transformed Hamiltonian
$\widetilde{H}%
_{0}=UH_{0}U^{\dag }$ takes the form
\begin{eqnarray}
\widetilde{H}_{0} &=&-\frac{1}{2}\nabla ^{2}+\frac{2l^{2}}{r^{2}}\hat{S}_{z}^{2}-%
\frac{2l}{r^{2}}\widetilde{L}_{z}\hat{S}_{z}+\frac{1}{2}r^{2}+2\Omega _{R}  \notag
\\
&&+\delta \hat{S}_{z}+\left( q-\Omega _{R}\right) \hat{S}_{z}^{2}+\sqrt{2}\Omega
_{R}\hat{S}_{x},  \label{C2}
\end{eqnarray}%
where $\widetilde{L}_{z}=-i\partial_{\phi}$ is the OAM operator in the rotating frame, and
the term proportional to $\widetilde{L}_{z}\hat{S}_{z}$ describes the SOAM coupling and plays a
critical role in our system.

\subsection{Single-particle energy spectrum}
We shall now find the energy spectrum determined by Hamiltonian $\widetilde{H}_0$. Obviously, $\widetilde{H}_{0}$ possesses rotational symmetry such that $%
\left[ \widetilde{L}_{z},\widetilde{H}_0\right] =0$. Therefore all the energy eigenstates can be labelled by two quantum numbers $|n, \widetilde{l}_{z} \rangle$ where $\widetilde{l}_{z}$ is the OAM quantum number, and $n$ may be regarded as the radial quantum number. Within a given $\widetilde{l}_{z}$ sector, the lowest energy eigenstate will be assigned $n=1$.

The eigenenergies and eigenstates can be easily found numerically by taking the ans\"atz \[ \widetilde{\Psi} = e^{i\widetilde{l}_z \phi}\, (u_1(r),\, u_0(r),\, u_{-1}(r) )^T \,.\] It is also easy to see that different spin states $|m \rangle$ ($m=-1$, 0, and 1) experience different effective potentials $V_m$ given by
\begin{eqnarray*}
V_{\pm 1} &=& \frac{\left( \widetilde{l}_{z}\mp 2l\right) ^{2}}{2r^{2}}+\Omega
_{R}+q\pm \delta +\frac{r^{2}}{2} \,,\\
V_{0} &=& \frac{\widetilde{l}_{z}^{2}}{2r^{2}}+2\Omega _{R}+\frac{r^{2}}{2}\,,
\end{eqnarray*}
where the term proportional to $1/r^2$ arises from the centrifugal barrier. Note that, in the lab frame, the OAM quantum number for spin state $|m \rangle$ is $\widetilde{l}_z \mp 2m l$. We will take experimentally interested LG modes $l=1$ and only focus on the two-photon resonant case $\delta=0$ in our following calculations. Furthermore, we choose $\Omega_0<0$ indicating that Raman beams are red single-photon detuning.

\begin{figure}[t]
\includegraphics[width=0.48\textwidth]{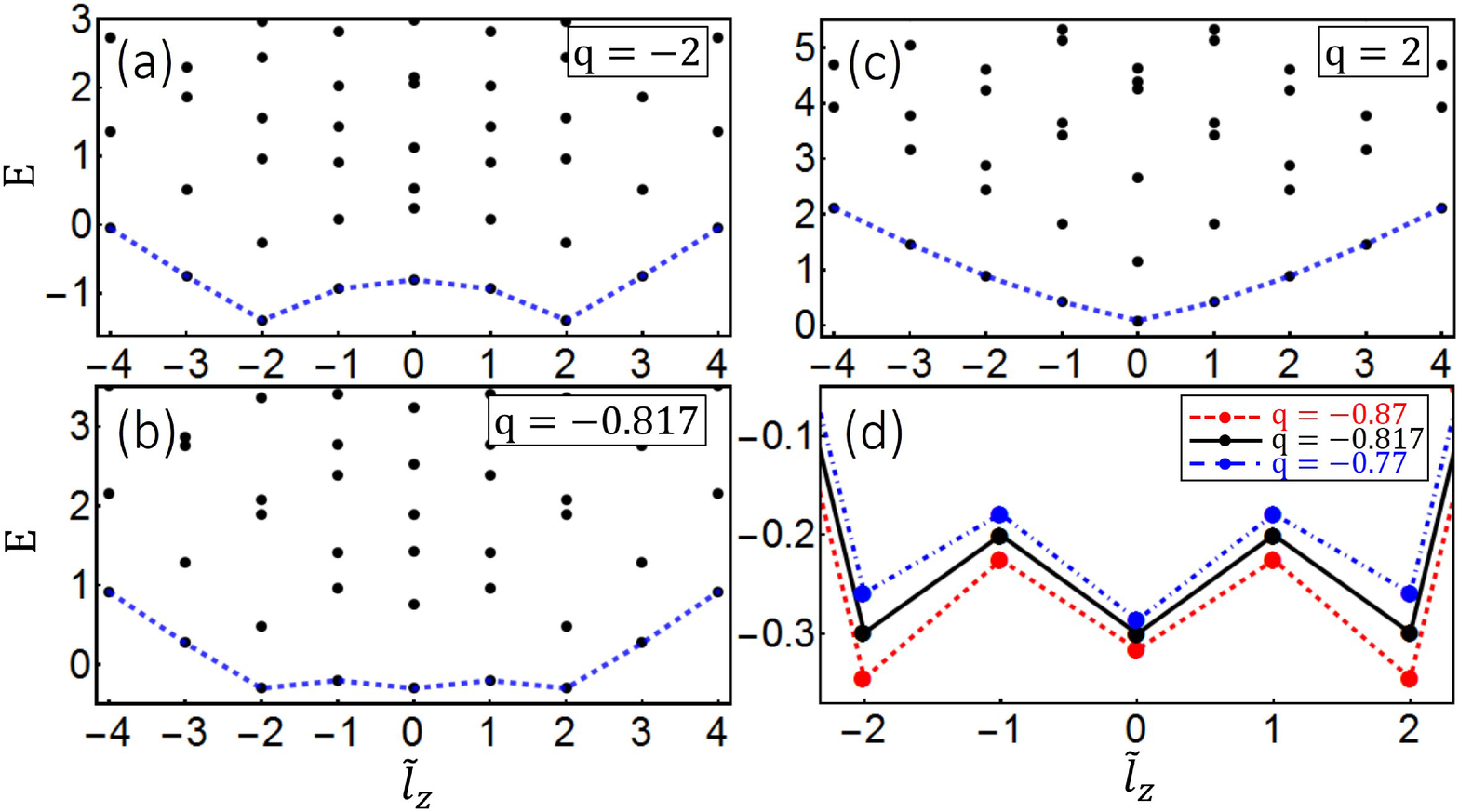}
\caption{(Color Online) The single-particle energy spectrum at $\Omega_{0}=-4$. (a), (b) and (c) correspond to $q = -2$, $q = -0.817$, and $q=2$, respectively, where the energy spectrum for the $n=1$ band (connected with dots) exhibit two, three and one minimum, respectively. (d) A detailed look at the $n=1$ band with three energy minima for three different values of $q$. At $q=-0.817$, all three minima are degenerate.}
\label{Fig2}
\end{figure}

Typical energy spectra for $\Omega_0=-4$ and several different quadratic Zeeman shift $q$ are presented in Fig.~\ref{Fig2}. One can easily see that all the spectra are symmetric about $\widetilde{l}_z=0$. This reflects an additional symmetry which is present only for $\delta=0$. This symmetry is associated with fact that the Hamiltonian $\widetilde{H}_{0}$ commutes with an
operator $\widetilde{T}=\widehat{A}%
\widehat{K}$, where $\widehat{K}$ denotes complex conjugation and
\begin{equation*}
\widehat{A}=%
\begin{pmatrix}
0 & 0 & 1 \\
0 & 1 & 0 \\
1 & 0 & 0%
\end{pmatrix} \,.
\end{equation*}%
It is straightforward to show that $\left\{
\widetilde{T},\widetilde{L}_{z}\right\} =0$ and $\left[ \widetilde{T},%
\widetilde{L}_{z}\right] =2\widetilde{T}\widetilde{L}_{z}$. As a result, $\widetilde{T}$ applying to an energy eigenstate $|n, \widetilde{l}_{z} \rangle$ changes the state to a degenerate eigenstate $|n, -\widetilde{l}_{z} \rangle$, i.e.,
\[ \widetilde{T}\left\vert n,
\widetilde{l}_{z}\right\rangle =\left\vert n,-\widetilde{l}_{z}\right\rangle \,,\]
which yields the symmetric spectrum.

One can also observe from Fig.~\ref{Fig2} that, according to the value of $q$, the spectrum may exhibit a single minimum at $\widetilde{l}_z=0$, two degenerate minima at $\widetilde{l}_z=\pm 2$, or three local minima at $\widetilde{l}_z =0, \pm 2$ in the $n=1$ band. Accordingly, we plot a `phase diagram' in Fig.~\ref{Fig1}(c). (For very large $|\Omega_0|$, the strong AC Stark shift confines the atoms along a thin ring and the energy spectrum is dominataed by one single minimum.) The yellow dashed line within the region with three minima corresponds to the case where all three minima are degenerate. For $\Omega_0=-4$, this occurs at $q\approx-0.817$, see Fig.~\ref{Fig2}(b). The wave functions of the three degenerate states are plotted in Fig.~\ref{Fig1}(d). If $q$ is slightly larger than this critical value, we have a global minimum at $\widetilde{l}_z=0$ and two local minima at $\widetilde{l}_z=\pm 2$; whereas for $q$ slightly smaller this critical value, global minima occur at $\widetilde{l}_z= \pm 2$, as shown in Fig.~\ref{Fig2}(d).

\section{weakly-interacting condensate}
In this section, we shall consider a weakly-interacting spin-1 condensate subject to the SOAM coupling, for which a mean-field treatment is appropriate. We work in a parameter regime where the single-particle spectrum exhibits three minima by taking $\Omega_0=-4$ and $q \in [-0.87, \,-0.77]$. This is the regime where the effects of the interatomic interaction can be most easily seen.

\subsection{Ground state phase diagram}

\begin{figure*}[t]
\includegraphics[width=0.98\textwidth]{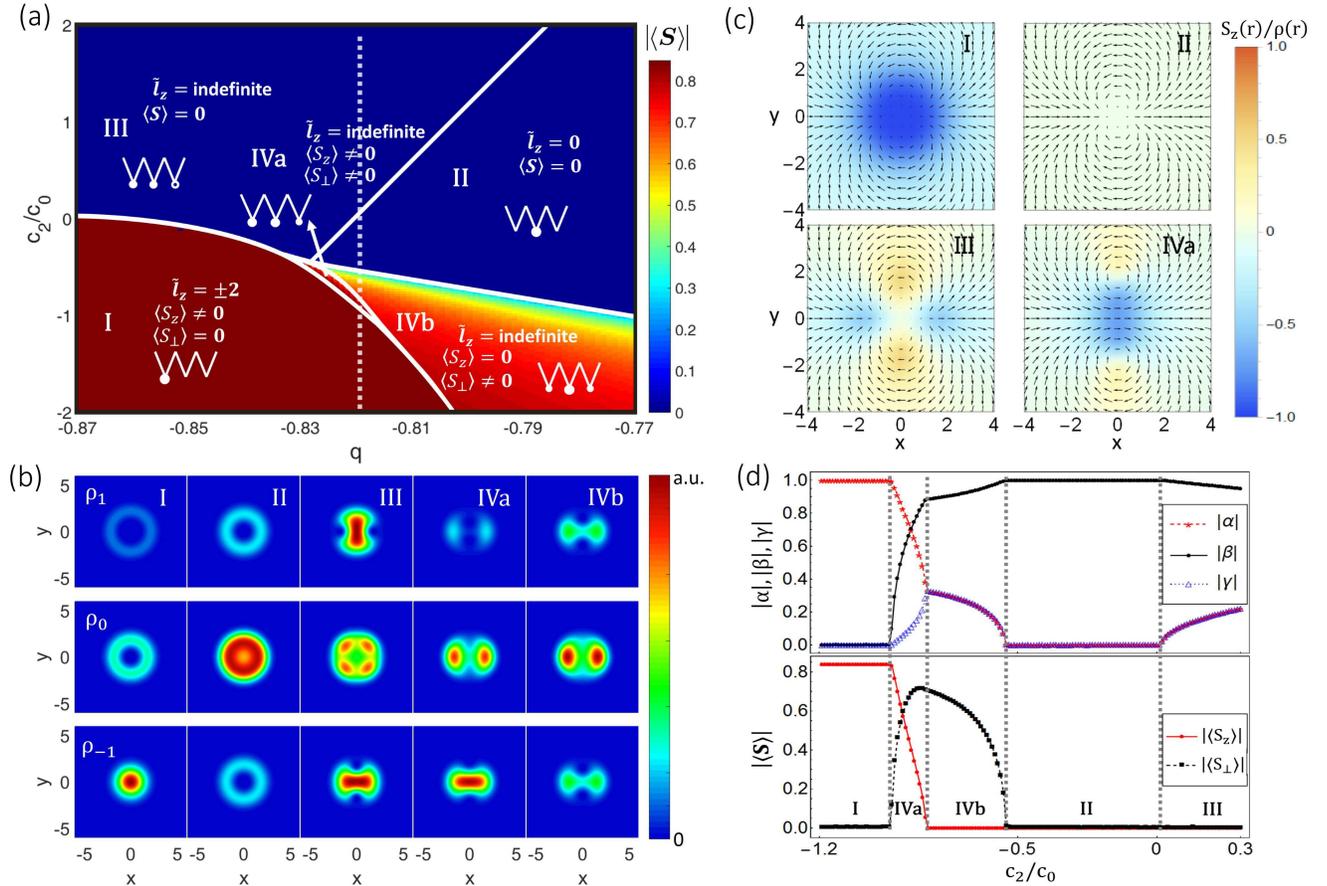}
\caption{(Color online) (a) Ground state phase diagram at $\Omega_{0} =-4$
and $c_{0}=1$. Background color represents the magnitude of the total spin $%
\left\vert\left\langle \bm{S} \right\rangle\right\vert$. Broken lines and disk-shaped markers show the relative weights of
$\protect\alpha$, $\protect\beta$ and $\protect\gamma$ in $\Psi$, where
hollow circle has a $\protect\pi$-phase difference with solid dots. (b)
Typical spin density distributions in each phase. (c) Normalized local spin texture $\bm{S}({\bf r})/\rho({\bf r})$ for different phases. The arrows represent transverse spin in the $xy$-plane and the background color represents longitudinal spin along the $z$-axis. Phase IVb (not shown) has similar transverse spin textures as Phase IVa but with $S_z({\bf r}) =0$.
(d) Top: Dependence of $\protect\alpha$ (red dashed line), $\protect\beta$ (black solid line) and $\protect\gamma$ (blue dotted line)
on $c_{2}$ at $q =-0.82$ corresponding to the white dashed vertical line in
(a). Bottom: Dependence of $\left\vert\left\langle \bm{S} \right\rangle%
\right\vert$ on $c_{2}$ at $q =-0.82$.}
\label{Fig3}
\end{figure*}

The interaction
Hamiltonian for a spin-1 condensate takes the form \cite{Kawaguchi,Stamper}
\begin{equation}
H_{\rm int}=\frac{1}{2} \int d^{2}r \,\left[ c_{0}\rho^{2}\left( {\bf r}\right) +c_{2}\bm{S}%
^{2}\left( {\bf r}\right) \right] \,, \label{C4}
\end{equation}%
where $\rho\left( {\bf r}\right)=\sum\nolimits_{m} \rho_m \left( {\bf r}\right)
=\sum\nolimits_{m}\left\vert \psi
_{m}\right\vert ^{2}$ is the total particle density
which obeys the normalization condition $\int d^{2}r \, \rho({\bf r})=1$, ${\bm S}({\bf r}) = \Psi^\dag \hat{\bf S} \Psi$ represents the local spin texture, $c_{0}$ is the
spin-independent two-body interaction satisfying $c_{0}>0$ (in our following calculation, we take $c_0=1$), and $c_{2}$ is
the spin-dependent interaction strength. In principle, $c_{0}$ and $c_{2}$ can be tuned by optical
Feshbach resonances \cite{Fatemi,Yan}. The total Hamiltonian of the condensate is given by \[ H = \int d^2 r \,
\Psi^\dagger H_{0}\Psi +H_{\rm int}\,.\] The ground state is obtained by minimizing the total energy. We study this problem using two different methods ---
the variational method and the fully numerical method based on the Gross-Pitaevskii (GP)
equations derived from Hamiltonian $H$. For the latter method, we performed imaginary-time evolution of the GP equations using the backward-Euler and Fourier pseudo-spectral discretizations toward
time and space, respectively \cite{Antoine2014}. The two methods produced results that are in excellent quantitative agreement. In the following, we just present our varitional calculation.

In the variational calculation, we assume that the condensate wave function is a linear superposition of the three lowest single-particle states $|n=1, \widetilde{l}_z=0, \,\pm 2 \rangle$:
\begin{equation}
\Psi \approx\alpha \left\vert 1,-2\right\rangle +\beta \left\vert
1,0\right\rangle +\gamma \left\vert 1,2\right\rangle \,,  \label{C5}
\end{equation}
where the complex amplitudes $\alpha $, $\beta $ and $\gamma$ satisfy the normalization condition $%
\left\vert\alpha\right\vert ^{2}+\left\vert\beta\right\vert
^{2}+\left\vert\gamma\right\vert ^{2}=1$. As it turns out, this is a very accurate approximation in the weakly-interacting regime that we are interested in. For discussion below, we define $\theta_\alpha$, $\theta_\beta$ and $\theta_\gamma$ to be the phase angles of these three amplitudes, respectively, i.e., $\alpha = |\alpha|\,e^{i \theta_\alpha}$, etc.

Taking $\alpha $, $\beta $ and $\gamma$ as variational parameters, we minimize the total energy to obtain the ground state. The total energy depends on these parameters in a rather complicated way. However, it depends on the three phase angles simply as $\cos (\theta_\alpha + \theta_\gamma - 2\theta_\beta)$. Hence, depending on the sign of the coefficient in front of it, the sum of the two angles $(\theta_\alpha + \theta_\gamma- 2\theta_\beta)$ can only take values 0 or $\pi$ (mod[$2\pi$]). This calculation allows us to distinguish several phases and we present the ground state phase diagram in Fig.~\ref{Fig3}(a). Typical spin density profiles $\rho_m({\bf r})$ and normalized local spin texture $\bm{S}({\bf r})/\rho({\bf r})$ for different phases are presented in Fig.~\ref{Fig3}(b) and (c), respectively. For a fixed $q=-0.82$, we plot the magnitude of the variational parameters and the total spin $\langle {\bm S} \rangle = \int d^2 r\,\bm{S}({\bf r})$ as functions of $c_2$ in Fig.~\ref{Fig3}(d). In the $c_2$-$q$ parameter space we explored, five distinct phases --- labelled as I, II, III, IVa
and IVb in the phase diagram --- are found. These phases result from the competition between the single-particle energies and mean-field interaction. The latter favors a ferromagnetic state with finite total spin when $c_2<0$, and an antiferromagnetic state with zero total spin when $c_2>0$. We describe the properties of these phases below.

Phase I --- This phase lies in the lower left corner of the parameter space. In this region, the single-particle ground state is two-fold degenerate with $\widetilde{l}_z = \pm 2$, see Fig.~\ref{Fig2}(d), and the interaction parameter $c_2<0$ favors a ferromagnetic state. As a result, the atoms condense into one of the single-particle ground states, and the many-body ground state is also two-fold degenerate and maintains rotational symmetry with definite OAM quantum number $\widetilde{l}_z = \pm 2$, corresponding to $\gamma=1$, $\alpha=\beta=0$ or $\alpha=1$, $\beta=\gamma=0$. The total spin is finite and points along the $z$-axis, i.e., $\langle S_z \rangle \neq 0$ and $\langle S_{\perp} \rangle=\sqrt{\langle S_x \rangle^2+ \langle S_y \rangle^2} =0$.

Phase II --- This phase lies in the upper right corner of the parameter space. In this region, the single-particle ground state is non-degenerate with $\widetilde{l}_z = 0$, and the interaction parameter $c_2>0$ favors an antiferromagnetic state. As a result, the atoms condense into the single-particle ground state, with $\beta=1$ and $\alpha=\gamma=0$, and a vanishing total spin $\langle {\bm S} \rangle =0$.

Phase III --- This phase lies in the upper left corner of the parameter space. Here $c_2>0$ favors an antiferromagnetic state, which results in a vanishing total spin $\langle {\bm S} \rangle =0$. In this phase, all three variational parameters $\alpha $, $\beta $ and $\gamma$ are nonzero with $|\alpha|=|\gamma|$ and
\begin{equation}
\theta_\alpha + \theta_\gamma- 2\theta_\beta=\pi \,. \label{III}
\end{equation}
As a result, the many-body ground state does not possess a definite value of $\widetilde{l}_z$ and spontaneously breaks the rotational symmetry. Typical spin density profiles are presented in the third column of Fig.~\ref{Fig3}(b), from which one can see that in the phase spin-1 and ($-1)$ components are immiscible. In this plot, we have chosen $\theta_\gamma=\pi$ and $\theta_\alpha=\theta_\beta=0$. A different choice of the angles under the constraint of Eq.~(\ref{III}) will result in a collective rotation of all three spin denisty profiles, which is a manifestation of the Goldstone mode resulting from the spontaneous symmetry breaking.

Phase IV --- This phase lies in the lower right corner of the parameter space. Here the single-particle ground state is non-degenerate with $\widetilde{l}_z = 0$, which can be regarded as an antiferromagnetic state. But the interaction $c_2<0$ is ferromagnetic. This competition again leads to all three variational parameters $\alpha $, $\beta $ and $\gamma$ to be nonzero and
\begin{equation}
\theta_\alpha + \theta_\gamma- 2\theta_\beta=0 \,. \label{IV}
\end{equation}
Similar to Phase III, the many-body ground state in Phase IV does not possess a definite value of $\widetilde{l}_z$ and spontaneously breaks the rotational symmetry. However, different from Phase III, here spin-1 and ($-1)$ components are miscible, as can be seen from the last two columns of Fig.~\ref{Fig2}(b), and the total spin $\langle {\bm S} \rangle$ does not vanish. The spin density profiles for Phase IV in Fig.~\ref{Fig2}(b) are plotted with $\theta_\alpha =\theta_\beta= \theta_\gamma=0$. Again, a different choice of phase angles under the constraint of Eq.~(\ref{IV}) will result in a collective rotation of all spin density profiles. In addition, Phase IV is the only phase that features a non-vanishing total transverse spin $\langle S_\perp \rangle \neq 0$. The local transverse spin forms a vortex-antivortex pair, see right lower corner of Fig.~\ref{Fig3}(c). Phase IV can be further decomposed into two subphases IVa and IVb. In IVa which only occupies a rather small parameter space, we have $\langle S_z \rangle \neq 0$. By contrast, in IVb, the amplitudes $|\alpha|=|\gamma|$ which, together with the phase angle  constraint in Eq.~(\ref{IV}), leads to $S_z({\bf r})=0$, i.e., the spin-1 and ($-1)$ components have identical density profiles, as can be seen in the last column of Fig.~\ref{Fig3}(b).

\subsection{Quench Dynamics}
After a detailed discussion of the ground state phase diagram, we now turn to the study of dynamics. In particular, we will examine how a sudden change of the quadratic Zeeman shift affects the system. Previous studies have shown that the quadratic Zeeman shift plays an important role in the quantum dynamics of a spinor condensate without spin-orbit coupling \cite{Kronj,Sadler}.

\begin{figure}[t]
\includegraphics[width=0.48\textwidth]{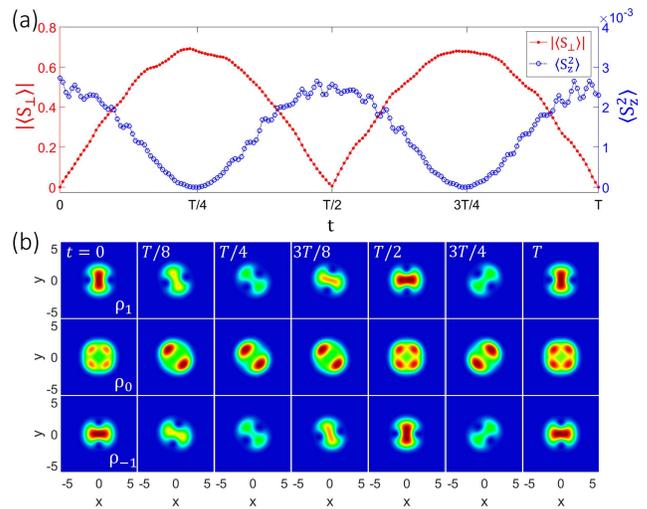}
\caption{(Color online) (a) Time evolution of total transverse spin magnitude $\left| \langle
S_{\perp}\rangle \right|$ (red solid curve with dots) and $\langle S_z^2 \rangle$ (blue dashed curve with hollow circles) after the quadratic Zeeman shift $q$ is quenched
from $-0.83$ to $-0.73$ at $t=0$. (b) Time evolution of the spin density profiles. The evolution is roughly periodic with a period of $T \approx 89$.
Here $c_2=1$.}
\label{Fig4}
\end{figure}

The time evolution of the system is depicted in Fig.~\ref{Fig4}. Initially we prepare the system in the ground state with $c_2=1$ and an initial quadratic Zeeman shift $q_i= -0.83$. This state belongs to Phase III. At $t=0$, we suddenly quench the quadratic Zeeman shift to a final value of $q_f= -0.73$ and the system starts to evolve. We solve the time evolution by numerically integrating the time-dependent GP equation with the help of real-time propagation method \cite{Antoine2015}.

In Fig.~\ref{Fig4}(a), we plot the evolution of the magnitude of the total transverse spin $|\langle S_\perp \rangle|$ which oscillates between the initial value of 0 and a maximum value, and $\langle S_z^2 \rangle = \int d^2r \,[\rho_1({\bf r}) - \rho_{-1} ({\bf r})]^2$ which oscillates between a finite initial value and 0. Note that $\langle S_z^2 \rangle = 0$ implies $\rho_1({\bf r}) = \rho_{-1} ({\bf r})$, i.e., identical density profiles for the spin-1 and ($-1$) components. The evolution of the spin density profiles is plotted in Fig.~\ref{Fig4}(b). It is quite evident that the evolution is periodic with a period $T \approx 89$. The system starts from an initial state that belongs to Phase III. At $t=T/4$, it evolves into a Phase IVb state. After another quarter period, it returns to Phase III, and the trend continues. Therefore the quench of $q$ induces a Rabi oscillation between two many-body quantum phases.

The above behavior obtained numerically can be understood using the variational ans\"atz Eq.~(\ref{C5}). The initial condensate wave function is very accurately described by Eq.~(\ref{C5}) with the amplitudes $|\alpha|=|\gamma|$, the phase angles $\theta_\alpha =\theta_\beta=0$, $\theta_\gamma=\pi$, and the single-particle states $|1, \widetilde{l}_z=\pm 2, 0 \rangle$ obtained at the initial quadratic Zeeman shift $q_i$. Immediately after the quench, we project the condensate wave function onto the single-particle states corresponding to the final quadratic Zeeman shift $q_f$. To a very good approximation, the condensate wave function is still dominated by the lowest band ($n=1$) single-particle states with $\widetilde{l}_z=\pm 2, 0$, i.e., we can write
\[ \Psi(t=0) \approx \alpha \left( 0\right) \left\vert
1,-2\right\rangle _{q_f}+\beta(0) \left\vert 1,0\right\rangle _{q_f}+\gamma (0)  \left\vert 1,2\right\rangle _{q_f} \,,\] where the amplitudes retain the relation $|\alpha(0)|=|\gamma(0)|$, $\theta_\alpha(0)=\theta_\beta(0)=0$ and $\theta_\gamma(0)=\pi$. In the ensuing time evolution, if we neglect the weak interaction energy, the condensate wave function will evolve according to
\[ \Psi(t) \approx \alpha \left( t\right) \left\vert
1,-2\right\rangle _{q_f}+\beta(t) \left\vert 1,0\right\rangle _{q_f}+\gamma (t)  \left\vert 1,2\right\rangle _{q_f} \,,\] with \[ \alpha(t)\! =\! \alpha(0) e^{-iE_{-2} t},\; \beta(t)\! =\! \beta(0) e^{-iE_{0} t},\; \gamma(t)\! =\!\gamma(0) e^{-iE_{2} t} \,,\]
where $E_{\widetilde{l}_z}$ is the single-particle energy for the state $|1,\widetilde{l}_z \rangle_{q_f}$. At $q_f=-0.73$, our calculation shows that $E_2=E_{-2}=E_0+\Delta$ with $\Delta\approx0.0501$. This leads to a periodic evolution of $\Psi(t)$ with period $T = 2\pi/\Delta \approx 125$, which has a little discrepancy with the numerically obtained $T \approx 89$. This discrepancy can mainly be attributed to the fact that we have neglected the interaction energy in our simple analysis, but the inclusion of the interaction energy would not affect the qualitative physics described here. Furthermore, the amplitudes will satisfy the condition $|\alpha(t)|=|\gamma(t)|$ and
\begin{eqnarray*}
\theta_\alpha(t)+\theta_\gamma(t) -2\theta_\beta(t) &=& \theta_\alpha(0)+\theta_\gamma(0) -2\theta_\beta(0) -2\Delta t\\ &=&\pi -2\Delta t \,.
\end{eqnarray*}
At $t=T/4=\pi/(2\Delta)$, we then have \[\theta_\alpha(T/4)+\theta_\gamma(T/4) -2\theta_\beta(T/4) =0 \,,\] and the condensate evolves into Phase IVb [see Eq.~(\ref{IV})], in agreement with the numerical calculation.

We have performed similar quenches starting from initial states within different phases. The many-body Rabi oscillation only occurs between Phases III and IVb. If the initial state belongs to either Phase I or II, the state is stable in the sense that it retains the rotational symmetry and the initial OAM quantum number $\widetilde{l}_z$.
If the initial state is within IVa, the post-quench dynamics looks rather complicated, and the system would not evolve into any other phases. Finally, we remark that, instead of a quench of $q$, a quench of the interaction strength $c_2$ can induce qualitatively similar dynamics. However, from an experimental point of view, quench of $q$ is much more feasible, and in fact has been realized in several laboratories \cite{Kronj,Sadler}.

\section{SUMMARY}

Motivated by previous experiments and recent theoretical studies of SOAM coupling in spin-1/2 condensate, we have presented a study of SOAM coupling in a spin-1 condensate. As we have shown, the enlarged spin degrees of freedom gives rise to much richer physics. Focusing on a regime where the single-particle energy spectrum exhibits a three-minima structure, we mapped out the ground state phase diagram where different phases possess distinct symmetry properties, spin density profiles and spin textures. We also investigated the dynamics induced by a sudden quench of the quadratic Zeeman shift, and found an interesting many-body Rabi oscillation between two of the phases. We have presented a variational analysis, along with a full numerical investigation, to provide a simple intuitive picture that underlies the main physics. The variational and the numerical calculations are in excellent agreement with each other.

\begin{acknowledgments}
YZ is supported by NSF of China under Grant Nos. 11234008 and
11474189, the National Basic Research Program of China (973 Program) under
Grant No. 2011CB921601, Program for Changjiang Scholars and Innovative
Research Team in University (PCSIRT)(No. IRT13076). HP
acknowledges support from US NSF and the Welch Foundation (Grant No.
C-1669).
\end{acknowledgments}

\end{document}